\documentclass{article}

\usepackage{amssymb}
\usepackage{graphicx}
\usepackage{amsmath,color,hyperref}
\usepackage{authblk} 

\title{Charge Transport at Atomic Scales in 1D-Semiconductors: A Quantum Statistical Model Allowing Rigorous Numerical Studies}

\author[1]{Roisin Braddell }
\author[2]{Jone Uria-Albizuri }
\author[3]{Jean-Bernard Bru}
\author[4]{Serafim Rodrigues}
\affil[1,3,4]{Basque Center for Applied Mathematics}
\affil[2,3]{Universidad del Pa\'{i}s Vasco}
\affil[3,4]{IKERBASQUE, Basque Foundation for Science}
\affil[1,2]{\small{\textrm{These authors contributed equally to the manuscript.}}}
\affil[2]{\small{\textrm{Corressponding Author: jone.uria@ehu.eus}}}

\begin{document}

\maketitle

\begin{abstract}

There has been a recent surge of interest in understanding charge
transport at atomic scales. The motivations are myriad, including understanding the conductance properties of peptides measured experimentally.  In this study, we propose a model of quantum statistical mechanics which aims to investigate the  transport properties of 1D-semiconductor at nanoscales. The model is a two-band Hamiltonian in which electrons are assumed to be quasi-free. It allows us to investigate the behaviour of current and quantum fluctuations under the influence of numerous parameters, showing the response with respect to
varying voltage, temperature and length. We  compute the current observable at each site and demonstrate the local behaviour generating the current.
\end{abstract}


\section{Introduction}

Recent advancements have increased interest in charge transport at atomic
scales. Growing demand for smaller electronic components and concurrent
improvements in experimental techniques, which allow the measurement of
currents across small scale conductors \cite{zhou2006direct,bjork2009donor,schmidt2009silicon} and even single peptides
\cite{zhang2019role}, have renewed interest in conductivity theory near the
atomic scale. At such scales, quantum effects can become relevant and indeed
behaviors which are not classical (for example, telegraph noise \cite
{zhang2019role}) have been experimentally measured. The question then
naturally arises on how to model such phenomena in the real world, for
example, for finite length conductors at finite temperatures. Indeed, it has
been shown that charge carrier hopping, an inherently quantum phenomenon, is
an important mechanism for charge transport in biological molecules \cite%
{genereux2010mechanisms,stubbe2003radical}.

Many quantum mechanical models investigate the behaviour of materials and
charge transport via electronic structure calculations at low temperatures,
based on different ansatz, approximations and numerical computations. Indeed, over the years, several theoretical and computational models have been developed to explain biological charge transfer (exchange of electrons occurring between an ionically conductive electrolyte and a protein in contact with the electrolyte) and charge transport (electrons flowing through a protein in the absence of electrolyte or electrolyte participation) \cite{Jochen2015}. The majority of these models are underpinned by the Marcus Theory of Electron Transfer \cite{ Marcus56B,Marcus56A} and extensions of it, know as Marcus-Hush theory \cite{Hush58}. This unifying theory expresses how electrons transition from an electron donor (D) to an acceptor (A) or sequences of D-A, mediated by oxidation-reduction (redox) centers (e.g. metal atoms, cofactors, amino acids/aromatic side chains, electrodes, etc) or electron sinks/sources (e.g. electrolytes). The primary goal of models built upon the Marcus-Hush theories have been to explain short-range electron transitions either via quantum-tunneling or hopping.

Examples of such models are, superexchange (SE), flickering resonance (FR), thermally-activated hopping (TH), diffusion-assisted hopping (DH) \cite{IngEtAl2018}. Other computational methods, include density functional theory calculations \cite{yan2010computational} or detailed molecular dynamics simulations \cite{sheu2002charge,sheu2015model}. In contrast, presently there is no unifying theory for long-range biological charge transport akin to the Marcus-Hush theory of electron transfer. Present approaches use sequential SE, FR, TH, DH steps, if for each step, the distance between redox centers are sufficiently short for effective tunneling ($\leq 20 \text{\AA}$ ), the energy levels of the redox centers are similar and there is strong coupling between electronic states \cite{BeratanEtAl2015}.
Long-range charge transport is hypothesized to occur through the formation of bands (e.g. mediated via formation of new bonds or crystalline structures) thus giving rise to electronic states that are delocalized within the entire peptide. In this case, it's conceivable to model free (conducting) electrons (or holes) as classical particles moving in continuous bands of delocalized electronic states, possibly leading to Ohm-like characteristics or nonlinear current-voltage as in  semiconductors. 

Thus, it is fundamental to develop unifying theories from first principles of quantum mechanics, which can under appropriate thermodynamical limits give rise to Ohm-like laws and as a by-product provide insights about possible mechanisms for long-range biological charge transport.

Fortunately, in a parallel literature, a community of mathematical physicists (including one of the authors) have been developing novel mathematically rigorous charge transport theories based on many-fermions systems on lattices \cite{OhmVI,Ansta-LDP,brupedraLR,OhmV,OhmIII,Ansta-LDP2} from first principles of quantum mechanics and the 2nd law of thermodynamics. These developments, for example, explain experiments showing that quantum effects vanish rapidly and macroscopic laws for charge transport emerge at length scales larger than a few nanometers. Specifically, Ohm's law survives for nanowires (silicon - Si doped with phosphorous - P) at 20 nm and even at low temperatures 4.2K) \cite{Ohm-exp, Ohm-exp2}.

The present manuscript attempts to exploit these novel theoretical frameworks but now focused towards explaining long-range charge transport in protein-ligand complexes that exhibit Ohmic characteristics superimposed with telegraph noise \cite{zhang2019role}. As a first attempt, we provide the simplest possible model of a finite length protein-ligand,  seen here as a $1D$-semiconductor. In particular, we assume the protein-ligand complex as a single molecule and we do not consider telegraph noise. Since we build upon these novel theoretical developments we will treat the underlying many-body system within the algebraic formulation of quantum mechanics applied to lattice fermion systems. Such systems can easily become computationally expensive or completely intractable and thus to make mathematical calculations amenable, we assume that the model is \emph{quasi-free}; meaning the inter-particle interaction is sufficiently weak. Then by invoking the rigorous mathematical results on our proposed lattice model we determine from the expectation values of microscopic current densities the classical Ohmic currents as well as possible semiconducting behaviors. This leads us with a preliminary insight into potential quantum effects. Moreover, by looking at the occupation numbers of the lattice for this $1D$-semiconductor, we observe directly the mechanism of charge transport.

\section{Description of the Model}\label{Description}

We model a $1D$-semiconductor via a two-band lattice - a conducting band and non-conducting band.  Spinless fermions are allowed to move along a conducting band (band $1$) or can get trapped in a second non-conducting band (band $0$). We write the model using the (usual) second quantization formalism.

\subsection{The Model at Equilibrium}

We first consider the Hamiltonian at equilibrium associated with our proposed two-band 1D lattice model. Omitting the spin of charge carriers, we thus use the one-particle Hilbert space defined by: 

\begin{equation}\label{hilbert space}
\mathfrak{h}:=\ell^{2} (\mathbb{Z} \times \{0,1\}) := \left \{(\psi (x))_{x\in\mathbb{Z}\times \{0,1\}}\subseteq \mathbb{C}:\sum_{x \in \mathbb{Z}\times\{0,1\}} \left \vert \psi \left( x\right) \right\vert ^{2}<\infty \right\}
\end{equation}%
while the fermionic Fock space is denoted by
\begin{equation}
\mathcal{F}:=\bigoplus_{n=0}^{\infty }P_{n}\mathfrak{h}^{\otimes n}
\label{fock space}
\end{equation}%
with $P_{n}$ being the projection to anti-symmetric functions of $\mathfrak{h}
$. Denote by $a_{x,\mathrm{b}}^{\ast }$ (resp. $a_{x,\mathrm{b}}$) the
creation (resp. annihilation) operator of a spinless fermion at $x\in
\mathbb{Z}$ in the band $\mathrm{b}\in \{0,1\}$ acting on the fermionic Fock space. We will now describe a general Hamiltonian for this particular system with parameters $\alpha_{x}, x \in \{\mathrm{p},\mathrm{r}\}$ which give the strength of the relevant terms for fermions in the protein (p) and reservoirs (r) respectively. How the strength of these terms is chosen is discussed in section \ref{ss:chooseparams}.

\subsubsection{Conducting Band}

Let $l\in \mathbb{N}$, $\epsilon _{\mathrm{p}}\geq 0$ and $\mu _{\mathrm{p}%
,1}\in \mathbb{R}$. We define the conducting band a 1D quantum system of length $2l\times \mathbf{a}$, $2l+1$ being the number of lattice sites and $\mathbf{a}$ (in $\mathrm{nm}$) being the lattice spacing, by the following Hamiltonian:

\begin{equation}
H_{\mathrm{p},1}:=\epsilon _{\mathrm{p}}\left( 2N_{\mathrm{p}%
,1}-\sum_{y,x\in \mathbb{Z}\cap \lbrack -l,l]:|x-y|=1}a_{y,1}^{\ast
}a_{x,1}\right) -\mu _{\mathrm{p},1}N_{\mathrm{p},1},
\label{conducting band hamiltonian 1D quantum system}
\end{equation}%
where
\begin{equation}
N_{\mathrm{p},1}:=\sum_{x=-l}^{l}a_{x,1}^{\ast }a_{x,1}
\end{equation}%
is the so-called particle number operator in band $1$. The above Hamiltonian has two well-identified parts:\ The first term in the
RHS of (\ref{conducting band hamiltonian 1D quantum system}) gives the kinetic energy
of fermions, this being the (second quantization of the) usual discrete Laplacian
with hopping strengths $\epsilon _{\mathrm{p}}\geq 0$ (in $\mathrm{eV}$).
The remaining part in the RHS of (\ref{conducting band hamiltonian 1D quantum system})
gives the basic energy level of the band $1$ inside the 1D quantum system, $\mu _{%
\mathrm{p},1}\in \mathbb{R}$ (in $\mathrm{eV}$) being the so-called chemical
potential associated with the conducting band.

\subsubsection{Insulating Band and Band Hopping}

Given $l\in \mathbb{N}$ and a chemical potential $\mu _{\mathrm{p},0}\in
\mathbb{R}$ (or Fermi energy), the Hamiltonian in the insulating band $0$ of
the quantum system (of length $2l\times \mathbf{a}$) is given as a basic energy level
$-\mu _{\mathrm{p},0}N_{0}$, where

\begin{equation}
N_{\mathrm{p},0}:=\sum_{x=-l}^{l}a_{x,0}^{\ast }a_{x,0}
\end{equation}%
is the particle number operator in the band $0$. In particular, no
hopping between lattice sites of the band $0$ is allowed. However, we
add a band-hopping term, allowing electrons to hop from one band to another
via the hopping strength $\gamma \geq 0$ (in $\mathrm{eV}$). Thus, we have the following Hamiltonian:
\begin{equation}
H_{\mathrm{p},0}:=-\mu _{\mathrm{p},0}N_{0}-\gamma \sum_{x=-l}^{l}\left(
a_{x,0}^{\ast }a_{x,1}+a_{x,1}^{\ast }a_{x,0}\right) .  \label{sdsdsdsds}
\end{equation}%

\subsubsection{The Fermion Reservoirs}

Given $l,L\in \mathbb{N}$ with $L\geq l$, $\epsilon _{\mathrm{r}}\geq 0$ and
$\mu _{\mathrm{r}}\in \mathbb{R}$, the system is assumed to be between two fermion reservoirs, the Hamiltonian of
which is given by%
\begin{equation}
H_{\mathrm{r}}:=\epsilon _{\mathrm{r}}\left( 2N_{\mathrm{r}}-\sum_{x,y\in
\mathbb{Z}\cap \left( \left[ -L,l-1\right] \cup \left[ l+1,L\right] \right)
:\left\vert x-y\right\vert =1}a_{y,1}^{\ast }a_{x,1}\right) -\mu _{\mathrm{r}%
}N_{\mathrm{r}}
\end{equation}%
where
\begin{equation}
N_{\mathrm{r}}:=\sum_{x=-L}^{l-1}a_{x,1}^{\ast
}a_{x,1}+\sum_{x=l+1}^{L}a_{x,1}^{\ast }a_{x,1}
\end{equation}%
is the particle number operator in the
reservoirs. We assume the reservoirs have a single band. Similar to the 1D quantum system, $\epsilon _{\mathrm{r}}$ and $%
\mu _{\mathrm{r}}$ (in $\mathrm{eV}$) give the hopping strength and chemical
potential inside the reservoirs, respectively. Note that the chemical
potentials in the left and right reservoirs (resp. in $\{-L,\ldots ,l-1\}$
and $\{l+1,\ldots ,L\}$) are the same in this case. Different chemical
potentials for each reservoir are possible, leading to similar, albeit
slightly more complex, behaviors. Thus, we refrain from considering this case in
our model to keep scientific discussions as simple as possible.

\subsubsection{The Full Hamiltonian}

Keeping in mind the electric connection between the 1D quantum system, the substrate
and the tip in \cite{zhang2019role}, the full Hamiltonian is equal to

\begin{equation}
H_{L}:=H_{\mathrm{p},1}+H_{\mathrm{p},0}+H_{\mathrm{r}}+H_{\mathrm{r-p}}
\label{full Hamiltonian}
\end{equation}%
where the Hamiltonian
\begin{equation*}
H_{\mathrm{r-p}}:=-\vartheta \left( a_{-l-1,1}^{\ast
}a_{-l,1}+a_{-l,1}^{\ast }a_{-l-1,1}+a_{l+1,1}^{\ast }a_{l,1}+a_{l,1}^{\ast
}a_{l+1,1}\right)
\end{equation*}%
allows fermions to hop between the reservoirs and the band $1$ of the 1D quantum system via the hopping strength $\vartheta \geq 0$ (in $\mathrm{eV}$).

For any $L,l\in \mathbb{N}$ with $L\geq l$, all Hamiltonians can be seen as
linear operators acting only on the restricted (fermionic) Fock space%
\begin{equation}
\mathcal{F}_{L}:=\bigoplus_{n=0}^{\infty }P_{n}\mathfrak{h}_{L}^{\otimes
n}\equiv \mathbb{C}^{2^{2\left( 2L+1\right) }}\subseteq \mathcal{F}
\label{Hilbert space L0}
\end{equation}%
constructed from the one-particle Hilbert space
\begin{equation}
\mathfrak{h}_{L}\doteq \left\{ \psi \in \mathfrak{h}:\forall x\in \mathbb{Z}%
\times \{0,1\}\backslash \left[ -L,L\right] ,\ \psi (x)=0\right\} \equiv
\ell ^{2}(\mathbb{Z}\cap \lbrack -L,L]\times \{0,1\}).
\label{Hilbert space L}
\end{equation}%

The dimension of the Fock space\ $\mathcal{F}_{L}$ grows exponentially with
respect to $L\in \mathbb{N}$, rapidly making a priori numerical computations expensive for $L\gg 1$. However, a key assumption of our proposed quantum model is its
quasi-free nature. This means that the many-fermion system can be entirely described within the one-particle Hilbert space, which is in this case equal to $%
\mathfrak{h}_{L}$. In particular, the numerical computations are therefore
done on a space of dimension $\mathrm{dim}\mathfrak{h}_{L}=2\left(
2L+1\right) $, instead of $\mathrm{dim}\mathcal{F}_{L}=2^{2\left(
2L+1\right) }$ as for general (possibly interacting) fermion systems.

\subsubsection{The Gibbs State}

In the algebraic formulation of quantum mechanics, a state $\rho$ is a
continuous linear functional on $\mathcal{B}\left( \mathcal{F}_{L}\right) $,
the Banach space of bounded linear operators on $\mathcal{F}_{L}$, that is
positive ($\rho \left( A\right) \geq 0$ if $A\geq 0$) and normalized ($\rho
\left( \mathbf{1}\right) =1$). If the system is at equilibrium at initial
time $t=0$, the equilibrium state is given by the Gibbs state associated
with the full Hamiltonian $H_{L}$, defined at temperature $\mathrm{T}>0$ (in
$\mathrm{K}$) and sufficiently large $L\in \mathbb{N}$ by
\begin{equation}
\rho \left( A\right) :=\frac{\mathrm{Trace}_{\mathcal{F}_{L}}\left( A\mathrm{%
e}^{-\left( \mathrm{k}_{B}\mathrm{T}\right) ^{-1}H_{L}}\right) }{\mathrm{%
Trace}_{\mathcal{F}_{L}}\left( \mathrm{e}^{-\left( \mathrm{k}_{B}\mathrm{T}%
\right) ^{-1}H_{L}}\right) },\qquad A\in \mathcal{B}\left( \mathcal{F}%
_{L}\right) ,  \label{Gibbs state}
\end{equation}%
where $\mathrm{k}_{B}$ is the Boltzmann constant (in $\mathrm{eV.K}^{-1}$).
As usual, it is convenient to use the parameter $\beta :=\left( \mathrm{k}%
_{B}\mathrm{T}\right) ^{-1}>0$, which is interpreted as the inverse
temperature of the system (in $\mathrm{eV}^{-1}$). This state can be studied
in the one-particle Hilbert space $\mathfrak{h}_{L}$ (\ref{Hilbert space L})
because it is a (gauge-invariant) quasi-free state, allowing us to greatly
simplify the complexity of the equations.

\subsection{The model driven by an electric potential}

\subsubsection{\textbf{Dynamics induced by electric potentials}}

Application of a voltage across the system results in a perturbed
Hamiltonian 

\begin{equation}
H_{L}\left( \eta \right) :=H_{L}+\eta E  \label{full Hamiltonian2}
\end{equation}%
with
\begin{equation*}
E:=-\sum_{x=-L}^{l-1}a_{x,1}^{\ast }a_{x,1}+\sum_{x=-l}^{l}\frac{x}{l}\left(
a_{x,1}^{\ast }a_{x,1}+a_{x,0}^{\ast }a_{x,0}\right)
+\sum_{x=l+1}^{L}a_{x,1}^{\ast }a_{x,1},
\end{equation*}%
$\eta \in \mathbb{R}$ (in $\mathrm{V}$) being a parameter controlling the
size of the voltage. Note that the electric potential difference is increases linearly across the protein. Note also that we take symmetric potential, meaning that the left reservoir (in $\{-L,\ldots ,l-1\}$) has its
chemical potential (or Fermi energy) shifted by $-\eta $, while the chemical
potential of the right reservoir (in $\{l+1,\ldots ,L\}$) is shifted by $%
+\eta $. The applied voltage on the quantum system is therefore $2\eta $. A
non-symmetric choice is of course possible, leading to similar, albeit
slightly more complex, dynamical behaviors. The symmetric choice is taken for the sake of simplicity.

In the algebraic formulation of quantum mechanics (cf. the Heisenberg
picture), the dynamics of the system is a continuous group $(\tau
_{t}^{(\eta )})_{t\in \mathbb{R}}$ defined on the finite-dimensional algebra
$\mathcal{B}(\mathcal{F}_{L})$ by
\begin{equation}
\tau _{t}^{(\eta )}\left( A\right) :=\mathrm{e}^{it\hbar ^{-1}H_{L}\left(
\eta \right) }A\mathrm{e}^{-it\hbar ^{-1}H_{L}\left( \eta \right) }\ ,\text{%
\qquad }A\in \mathcal{B}(\mathcal{F}_{L}),  \label{dynamics}
\end{equation}%
at any time $t\in \mathbb{R}$ (in $\mathrm{s}$). In particular, a physical
quantity represented by an observable $A$ becomes time-dependent. The
expectation value of this physical quantity value is given by the real
number $\rho (\tau _{t}^{(\eta )}(A))$, its variance by $\rho (\tau
_{t}^{(\eta )}(A)^{2}-\rho (\tau _{t}(A))^{2})$, and so on, since the initial
state of the system ($t=0$) is given by the Gibbs state $\rho $ defined by (%
\ref{Gibbs state}).

Like the Gibbs state, the resulting dynamics are quasi-free and it can thus be studied in the one-particle Hilbert space $\mathfrak{h}_{L}$ (\ref{Hilbert space L}%
). This feature allows us to greatly simplify the complexity of the
equations. 

\subsubsection{Current Observables}

The definition of current observables is model dependent in general. To
compute them, it suffices to consider the discrete continuity equation (in terms of observables) for the fermion-density observable

\begin{equation}
n_{x}(t)\doteq \tau _{t}^{(\eta )}(a_{x,b}^{\ast }a_{x,b})
\end{equation}%
at lattice site $x\in \{-L,\ldots ,L\}\mathbb{\ }$and time $t\in \mathbb{R}$
in the band $b\in \{0,1\}$:%
\begin{equation}
\partial _{t}n_{x,b}(t)=\tau _{t}^{(\eta )}\left( i\hbar ^{-1}\left[
H_{L}\left( \eta \right) ,a_{x,b}^{\ast }a_{x,b}\right] \right)
\end{equation}%
where $[A,B]\doteq AB-BA$ is the usual commutator. For any fixed $x\in
\{-L,\ldots ,L\}$ and $b=1$, one straightforwardly computes from the
Canonical Anticommutation Relations (CAR) that 

\begin{eqnarray}
\partial _{t}n_{x,1}(t) &=&\gamma \tau _{t}^{(\eta
)}(I_{(x,x)}^{(0)})+\epsilon _{\mathrm{p}}\sum_{y\in \mathbb{Z}\cap \left[
-l,l\right] :\left\vert x-y\right\vert =1}\tau _{t}^{(\eta
)}(I_{(x,y)}^{(1)})  \label{current observable2} \\
&&+\vartheta \tau_{t}^{(\eta)}\left(\delta_{x,-l} I_{(-l-1, x)}^{(1)}+\delta_{x, l} I_{(l+1, x)}^{(1)}+\delta_{x,-l-1} I_{(x,-l)}^{(1)}+\delta_{x, l+1} I_{(x, l)}^{(1)}\right) \notag \\
&&+\epsilon _{\mathrm{r}}\sum_{y\in \mathbb{Z}\cap \left( \left[ -L,l-1%
\right] \cup \left[ l+1,L\right] \right) :\left\vert x-y\right\vert =1}\tau
_{t}^{(\eta )}(I_{(x,y)}^{(1)}),  \notag
\end{eqnarray}%
($L>l>1$) where, for any $x,y\in \{-L,\ldots ,L\}$ and $b\in \{0,1\}$,
\begin{equation}
I_{(x,y)}^{(b)}:=i\hbar ^{-1}\left( a_{x,1}^{\ast }a_{y,b}-a_{y,b}^{\ast
}a_{x,1}\right) .  \label{current observables}
\end{equation}%

The positive signs in the right-hand side of (\ref{current
observable2}) results from the fact that the particles are positively charged, $%
I_{(x,y)}$ being the observable related to the flow of positively charged,
zero-spin, particles from the lattice site $x$ to the lattice site $y$.
Negatively charged particles can be considered in the same way. In
fact, there is no experimental data on the sign of the charge carriers in
\cite{zhang2019role}, even if it is believed that the proximity of bands
associated with oxidizable amino acid residues to the metal Fermi energy
suggests that hole transport is more likely.

We calculate the current inside the system, that is, we calculate the second term in the RHS of equation (\ref{current observables}). Therefore, the current density observable in $\{-l,\ldots ,l\}$ produced by the electric potential
difference $2\eta \geq 0$  at any time $t\in \mathbb{R}$ is thus equal to
\begin{equation}\label{current density observable}
\mathbb{J}\left( t,\eta \right) \doteq \frac{\epsilon q_{e}}{2l}%
\sum_{x=-l}^{l-1}\{\tau _{t}^{(\eta )}(I_{\left( x+1,x\right) }^{(1)})-\tau
_{t}^{(0)}(I_{\left( x+1,x\right) }^{(1)})\}
\end{equation}%
with $q_{e}$ being the charge (in $\mathrm{C}$) of the fermionic charge
carrier (electron or hole). Note that we remove the possible free current on
the systems when there is no electric potential, even if one can verify in this specific case that it is always zero at equilibrium. Observe also that we do not consider (i)
currents in the left and right fermion reservoirs, (ii) contact currents
from the reservoirs to the 1D quantum system (in $\{-l,\ldots ,l\}$) and
(iii) currents from the conducting band $1$ to the trapping band $0$. These
currents can easily be deduced from Equation (\ref{current observable2}):
for (i)-(iii), see the terms in (\ref{current observable2}) with $\epsilon _{%
\mathrm{r}}$, $\vartheta$ and $\gamma$ respectively.

\subsubsection{Current Expectations and Fluctuations}

Expectation values of all currents are obtained by applying the state of the
system. For instance, if the initial state of the system is the Gibbs state $%
\rho $ defined by (\ref{Gibbs state}), the expectation of the current
density (in $\mathrm{A}$) produced by the electric potential difference $%
2\eta \geq 0$ (in $\mathrm{V}$) in $\{-l,\ldots ,l\}$ at any time $t\in
\mathbb{R}$ (in $\mathrm{s}$) and temperature $\mathrm{T}>0$ (in $\mathrm{K}$%
) equals
\begin{equation}
\mathbb{E}\left( \mathbb{J}\left( t,\eta \right) \right) :=\rho \left(
\mathbb{J}\left( t,\eta \right) \right) .
\label{current density observable2}
\end{equation}

Quantum fluctuations of an observable $A$ are naturally given by the variance
\begin{equation}
\text{Var}(A)=\mathbb{E}[A^{2}]-\mathbb{E}[A]^{2}.
\label{current density observable3}
\end{equation}%

Applied to the current observables, this leads to the current variance. More
generally, all moments associated with current observables can be defined in
a similar way, according to usual probability theory. We refrain from considering
higher moments than the variance in the sequel\ in order to keep things as
simple as possible. This is however an important question to be studied in
future research works in the context of the telegraph noise.
Indeed, as shown in \cite[Fig. S7]{zhang2019role}, the telegraph noise is
characterized by a bimodal distribution of currents, which could be investigated by 
some statistical methods involving 3rd and 4th order moments (such as the
kurtosis and skewness).

Note finally that the current density is an average of currents over the line $\{-l,\ldots ,l\}$ and it should converge rapidly to a deterministic value, as $l \to \infty$. Similar to the central limit theorem in standard probability theory, one could study the rescaled variance 

\begin{equation*}
2l\text{Var}\left( \mathbb{J}\left( t,\eta \right) \right) =\frac{1}{2l}%
\left( \mathbb{E}\left( \left( 2l\mathbb{J}\left( t,\eta \right) \right)
^{2}\right) -\mathbb{E}\left( 2l\mathbb{J}\left( t,\eta \right) \right)
^{2}\right),
\end{equation*}
which should converge to a fixed value. As shown in \cite{Ansta-LDP2} for a one-band model, this quantity should be directly related with the rate of convergence of the current density, as $l \to \infty$. This is however not studied in detail here.

\subsection{Choosing the Parameters of the System\label{ss:chooseparams}}

We subsequently fine tune the proposed model by appropriately choosing a physiologically suitable parameter set.  However, because of the novelty of the protein-ligand complexes experiment outlined in
\cite{zhang2019role}, the associated microscopic parameters and data are still scant. Nevertheless, our aim is to provide a general theoretical framework, where particular model instances can first lead to qualitative agreement and future improvements to quantitative and complete explanation of this phenomenon. Consequently, we choose parameters that are physically plausible and with the correct order of magnitude. 

In general, we will assume ``room temperature", i.e., $\mathrm{T}\approx 300\mathrm{\ K}$ and this determines the value of $\beta :=\left( \mathrm{k}_{B}\mathrm{T}\right)
^{-1}\approx 38.7$ $\mathrm{eV}^{-1}$. The lattice length, $l$, will be set in accordance to the peptides considered in \cite{zhang2019role}. However, we
first need to fix the lattice spacing $\mathbf{a}$ of our model. In
crystals, the lattice spacing is of the order of a few angstroms, and it is
natural to assume roughly the same order of magnitude inside molecules. In
our prototypical example, we set\footnote{%
E.g., in pure silicone (Si), it is about $0.2\ \mathrm{nm}$ \cite{Metrologia}. In
the 2D lattice formed by $\mathrm{CuO}_{4}$ in the cuprate superconductor $%
\mathrm{La}_{2}\mathrm{CuO}_{4}$, the lattice spacing of the oxygen ions is $%
0.2672\ \mathrm{nm}$ \cite[Sect. 6.3.1]{unitcoherence}, corresponding to a
lattice spacing of the copper ions equal to $0.3779\ \mathrm{nm}$.} $\mathbf{%
a}\simeq 0.3\ \mathrm{nm}$. Following \cite{zhang2019role}, we consider a 1D quantum system of several nanometers,
more precisely of a size between $2\ \mathrm{nm}$ and $10\ \mathrm{nm}$ or
even a little more. This corresponds to a 1D quantum system of length $2l$ between about $6.66$ and $33.33$, i.e., $l\in \lbrack 3,17]$. For
typical experiments we choose $l=5$ (i.e., $3\ \mathrm{nm}$) and explore the
effect of varying the length of the 1D quantum system.
Concerning the length\ of the reservoirs, they have to be as large as
possible in the sense that $L\gg l$ and our choices are constrained by
the computational tractability. Moreover, we
choose $L\gg l$ in such a way as to eliminate artificial \textquotedblleft
finite size\textquotedblright\ effects, as discussed in Section \ref{Finite
Size Effects}.

Appropriate ranges of $\epsilon _{\mathrm{p}}$ and $\epsilon _{\mathrm{r}}$,
that is the hopping strength of a fermion in the 1D quantum system (in $%
\{-l,\ldots ,l\}$, band $1$) and fermion reservoirs respectively, can be
derived from the lattice spacing $\mathbf{a}$ and the effective mass of
charge carriers $m^{\ast }\simeq Cm_{e}$, where $m_{e}$ is the electron
mass. If $C=1$, then a general hopping strength equals
\begin{equation*}
\epsilon :=\frac{\hbar ^{2}}{m^{\ast }\mathbf{a}^{2}}=\frac{\hbar ^{2}}{Cm_{e}\mathbf{a}^{2}}\simeq 0.85\ \mathrm{eV}.
\end{equation*}%

We set $\epsilon _{\mathrm{r}}=0.85$ inside the reservoirs and take epsilon slightly smaller inside the protein, $\epsilon _{\mathrm{p}}=0.65\text{ }\mathrm{eV}$ ($C \simeq 1.3$). In real systems, the effective mass of an electron is usually larger than  the electron mass, but since we have no concrete information at our disposal we only make the reservoir more conducting than the conducting band of the 1D quantum system. We note that changes to epsilon on the magnitude of  $0.1-0.2\text{ }\mathrm{eV}$ do not have a significant effect on the behavior of the system, unlike changes to the $\gamma$ parameter. 

Similarly, the hopping strength $\vartheta $ controlling the hopping
strength between the 1D quantum system and fermionic reservoirs should be of
the same order. For simplicity, we choose $\vartheta =(\epsilon _{\mathrm{p}}+\epsilon _{\mathrm{r}})/2$ , meaning a very good hopping contact between the 1D quantum system and the two reservoirs. In real systems, this variable could be modified to encode bad connections between reservoirs and the 1D quantum system.

\begin{figure}
    \centering
    \includegraphics[scale=0.45]{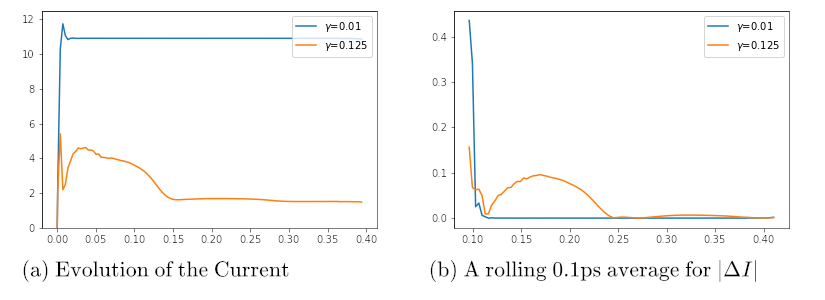}
    \caption{Comparison of current evolution and absolute difference of the current between timesteps of $0.01 \text{ } \mathrm{ps}$ for the conductor type model ($\protect\gamma=0.01$) and semi-conductor type model ($\protect\gamma=0.125$) for $\eta=0.1$. For $\protect\gamma=0.01$ the time taken to reach the stationary regime is $0.1\text{ } \mathrm{ps}$, for $\protect\gamma=0.125$ the time taken is approximately $0.25 \text{ } \mathrm{ps}$.}
    \label{conductorvssemiconductor}
\end{figure}

The choice of $\gamma $ fundamentally alters the nature of the model, with
the 1D quantum system acting as a conductor or exhibiting semi-conductor like
behaviors depending on the choice of $\gamma $. We explore both situations.

Finally, note that the voltage applied in the experiments described in \cite%
{zhang2019role} is of the order of one tenth of volts. For instance, for $%
0.1\ \mathrm{V}$ applied on the 1D quantum system, we are still in the linear response
regime without any telegraph noise, see \cite[Fig. 1]{zhang2019role}. Naively, this
looks coherent with the energy scale used here, which is of the order of tenth of electronvolts.

\section{Methods}

Using the quasi-free property of the model one can calculate the evolution of
the expectation value of the current density. For quasi-free equilibrium
states and dynamics, the phase space can be reduced from the Fock space,
which is of dimension $2^{2(2L+1)}$, to a one-particle phase space of
dimension $2(2L+1)$. This is achieved by replacing the Hamiltonian with an
equivalent one-particle Hamiltonian which defines dynamics on the
one-particle Hilbert space $\mathfrak{h}_{L}$ (\ref{Hilbert space L}). This allows for an efficient computation of quantities (\ref{current density observable2}) and  (\ref{current density observable3}).

The model was written in Python and is largely built with the NumPy package. The code is well-suited for calculating higher-order powers of observables (in this case, the current observable) which are  needed to investigate the statistical properties of these observables. The limitation here is increasing computational complexity (see for instance Supplementary material section \ref{times}) which could rapidly become a problem with the extension of this method to higher dimensions. However, this approach is well-adapted to nanometric quantum systems as found, for example, in biological systems. The calculations of charge transport were performed on a MacBook Pro with a 2,4 GHz Quad-Core Intel Core i5-processor.

\section{Results}

The nature of the model is altered by hopping terms in (\ref{sdsdsdsds})
between both bands, the strength of which is the parameter $\gamma $. For $%
\gamma=0 $, or $\gamma$ close to $0$ the system behaves as a 1D-conductor. The expectation
value (\ref{current density observable2}) of the current is directly
proportional to the applied voltage until it saturates, by reaching a
maximum value. As $\gamma$ increases the model starts to exhibit semi-conductor like behavior, with low current until a threshold value of voltage at which the conducting band begins to exhibit partially filled charge-carrier states. After the threshold voltage, the
current (\ref{current density observable2}) increases in an approximately
linear manner until it saturates.

\subsection{Reservoirs and Finite Size Effects\label{Finite Size Effects}}

\begin{figure}
    \centering
    \includegraphics[scale=0.45]{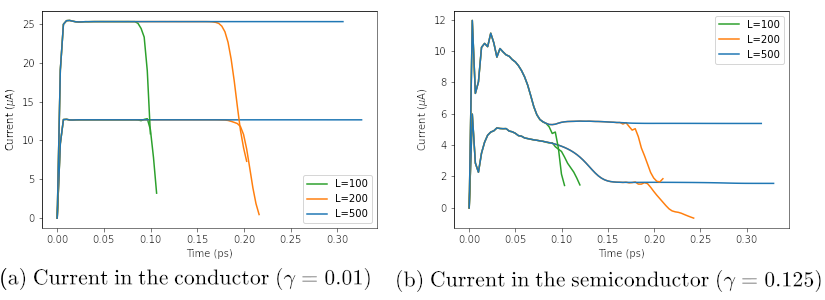}
    \caption{Evolution of the current for different lengths of the
charge-carrier reservoirs for $\eta\in \{0.1,0.2\}$ and a variety of different reservoir lengths showing the finite duration of the stationary regime and subsequent current collapse.}
    \label{currentcollapse}
\end{figure}

\begin{figure}
    \centering
    \includegraphics[scale=0.45]{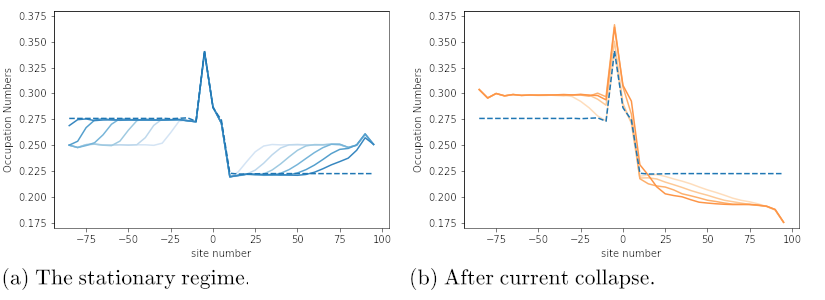}
    \caption{Occupation number evolution for the conductor case ($\gamma=0.01$) during stationary regime (a) and after current collapse (b). Larger times are encoded via bolder blue/orange lines. The dashed lines represent the values for the limit state, i.e., the Gibbs state when the full system is in presence of an electric potential $\eta=0.1$.}
    \label{occupationnumbercollapse}
\end{figure}

Numerical simulations were performed for numerous different parameters. For a given set of parameters, we calculate the mean current by disregarding transients and specifically averaging the current over the course of $0.1\ \mathrm{ps}$ in the \emph{steady regime}, i.e., in the time interval in which the current change is negligible. For practical purposes, we say the current change is negligible when it deviates by less than $10^{-3} \mu A$ over $0.1\ \mathrm{ps}$. See Fig. \ref{conductorvssemiconductor}(b). In practice, the steady regime  exists for a finite interval of times depending on the sizes of the reservoirs, which are directly tuned by the parameter $L\in \mathbb{N}$. This can be seen in Fig. \ref{currentcollapse}, where the length of the steady regime always increases with respect to $L$ and would be stationary for all sufficiently large times in the limit $L\rightarrow \infty $. We do not provide here a mathematical proof of this conjecture, but this feature was present in all our numerical computations.

In fact, our numerical computations inevitably induce finite size effects.
This can be well understood via the depletion of charge carriers in the
finite reservoirs, which yield a current collapsing as seen in Fig. \ref{currentcollapse} for finite $L$ and sufficiently large times. This depletion is
explicitly demonstrated in Fig \ref{occupationnumbercollapse} which gives the
fermionic occupation number
\begin{equation}
\mathbb{E}\left( \tau _{t}^{(\eta )}\left( a_{x,1}^{\ast }a_{x,1}\right)
\right) :=\rho \left( \tau _{t}^{(\eta )}\left( a_{x,1}^{\ast
}a_{x,1}\right) \right)
\label{fermion occupation number}
\end{equation}%
at time $t\geq 0$ in each lattice site $x\in \{-L,\ldots ,L\}$ in the band $1
$. One can observe that the depletion of the occupation number is directly correlated to the collapsing of the current, as expected with a naive classical viewpoint.

\subsection{Emergence of Semi-Conductors}

Although the experiments performed in \cite{zhang2019role} show that a set of protein-ligand complexes display Ohm-like conductivity (i.e. possibly via the formation of new molecular bonds that give rise to delocalized electronic states akin to conductors) we also predict formation of semi-conductor type characteristics. Although this has not been experimentally observed so far, our model predicts the existence of protein-ligand complexes with such a behavior, as anticipated for this kind of model. This is achieved by first noticing that the two bands in the model are linked with each other via the terms

\begin{equation*}
-\gamma \left( a_{x,0}^{\ast }a_{x,1}+a_{x,1}^{\ast }a_{x,0}\right) ,\qquad
\gamma \geq 0,\ x\in \left\{ -l,\ldots ,l\right\} ,
\end{equation*}%

for all lattice sites inside the system of lattice number $2l$ (corresponding to a 1D quantum system of length $2l\times \mathbf{a}$, $\mathbf{a}$ being the lattice spacing in $\mathrm{nm}$),
see (\ref{sdsdsdsds}). The parameter $\gamma $ allows us to control the
nature of the conductor.\

\begin{figure}
    \centering
    \includegraphics[scale=0.45]{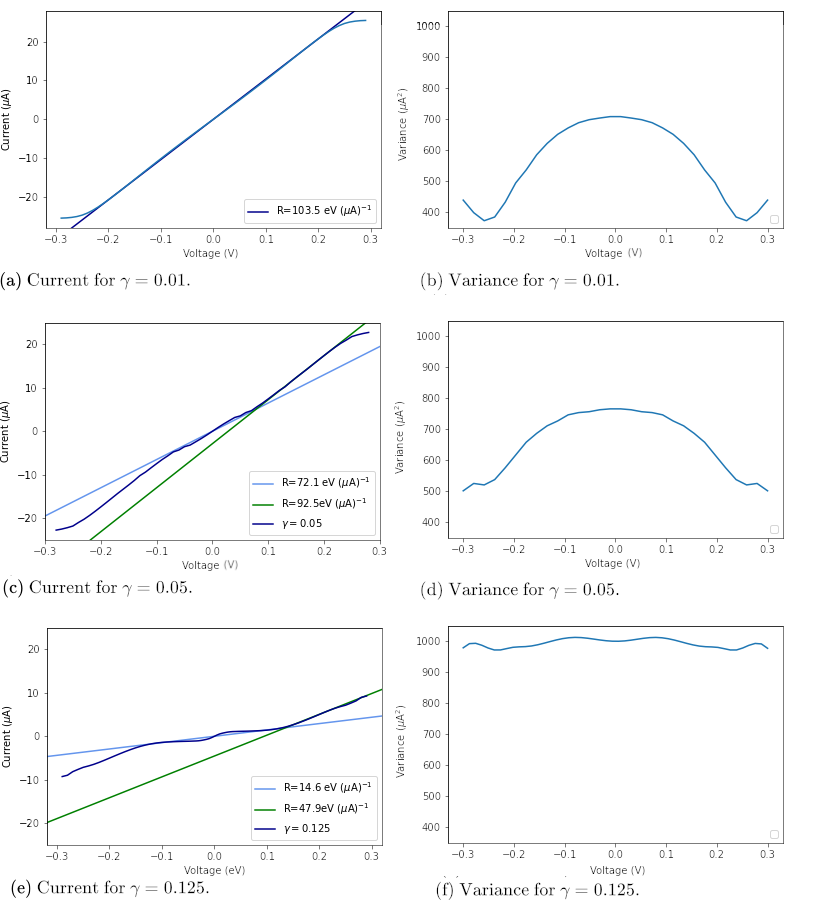}
    \vspace*{2mm}
    \caption{Plots (a), (b) and (c) show the expected value of current vs the parameter $\eta$ (representing the electric potential) in the conducting ($\protect\gamma=0.01$), intermediate ($\protect\gamma=0.05$) and semiconducting ($\protect\gamma=0.125$) regimes respectively. Plots (b), (d) and (f) give the corresponding expected values of the variance.  As $\protect\gamma$ increases further the 1D quantum system begins to exhibit semi-conductor like behavior, with high-resisitivty at lower voltages and a transition to conductor like behavior at a threshold voltage. $R$ in figures (a), (d) and (e) gives the slope of the best fit straight line.}
    \label{etatests}
\end{figure}

For $\gamma=0$, the two bands are disconnected and the system
behaves as a regular conductor. In particular, the current increases
linearly with voltage for sufficiently small electric potentials. This behavior is seen for $\gamma \ll 1$ and the same behavior holds true for small $\gamma \neq 0$, see, e.g., Fig. \ref{etatests}(a). All these
phenomena are of course expected and can in fact be mathematically showed in great
generality, see, e.g., \cite{OhmV,OhmVI}, even in the limit of infinitely large $L\gg l\gg 1$.

\begin{figure}
    \centering
    \includegraphics[scale=0.45]{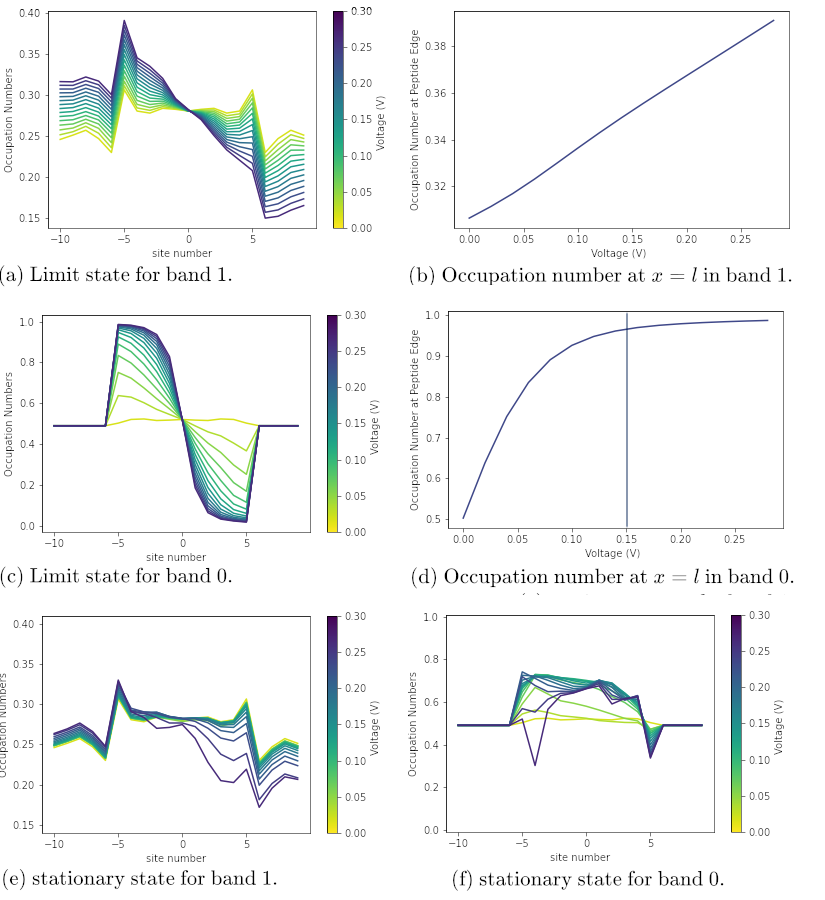}
    \caption{Plots (a) and (c) show the fermionic occupation numbers in the limit (Gibbs) state at different values of electric potential $\eta$ for the semiconductor ($\gamma=0.125$) in band $1$ (conducting band) and band $0$ (insulating band) respectively. Plots (b) and (d) show the size of the occupation number peak in at the edge of the protein (i.e. the site $x=l$) as a function of $\eta$ in band $1$ and band $0$ respectively. At lower voltages the second band has an effect on the current. As the voltage increases the fermionic density in the insulting band starts to saturate at $\eta \approx 0.15$, at which point the conducting behavior dominates. Plots (e) and (f) show the expected values of the occupation numbers during the stationary regime in band $1$ and band $0$ for different values of $\eta$.}
    \label{occupationnumbereta1}
\end{figure}

For $\gamma \neq 0$ the linear relationship between current and voltage progressively disappears as the average resistivity increases at lower voltages. The origin of this behavior can be studied by looking at the occupation numbers of the different lattice sites for the limit Gibbs state at different values of $\eta$, see Figure \ref{occupationnumbereta1}. For non-zero gamma as  $\eta$ increases the occupation numbers for the limit Gibbs state in the non-conducting band also increase. As this insulating band becomes saturated at the $x={-l,l}$ lattice-sites, the 1D quantum system once again starts to act as a conductor. For sufficiently strong $\gamma$, this phenomenon starts to be visible in terms of the charge transport within the (conducting) band 1: 

In Figure \ref{occupationnumbereta1}(b), we anticipate a change of current response at a saturation voltage of approximately $0.15 \mathrm{V}$, which is precisely the case, as shown in \ref{etatests}. In fact, at $\gamma \approx 0.1$ the system starts to exhibit this semi conductor-like
behavior, see Figures \ref{etatests}(c) and \ref{etatests}(e). This behavior is concomitant with the current fluctuations given in Figs. %
\ref{etatests}(b)--\ref{etatests}(f): The high electric resistance for small voltages is associated with an increase of the current fluctuations until some saturation regime (corresponding to the voltage threshold) from which the fluctuations decrease again, as in the conductor case. Note that the current fluctuations once again increase in any case for sufficiently large voltage beyond the linear (resp. piece wise linear) regime in the conductor (resp. semi-conductor) case.  

\begin{figure}
    \centering
    \includegraphics[scale=0.45]{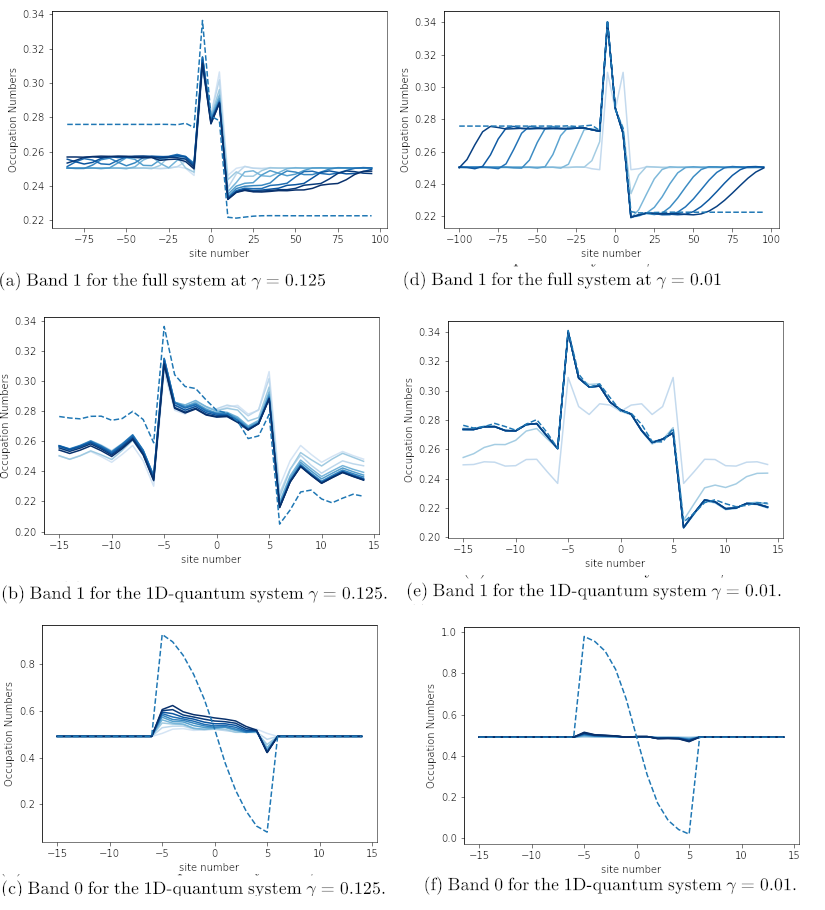}
    \caption{Comparison of the expected value of the fermionic occupation number at $\eta=0.1$ at the lattice sites for the semi-conductor (plots (a), (c) and (e)) and conductor ($\gamma=0.01$, plots (b), (d) and (f)). Larger times are encoded via bolder blue lines. The dashed lines represent the values the expected limit Gibbs state, when the full system is in presence of an electric potential. The 1D quantum system lies between the sites $-5$ and $5$.}
    \label{occupationnumbertime1}
\end{figure}

\begin{figure}
    \centering
    \includegraphics[scale=0.3]{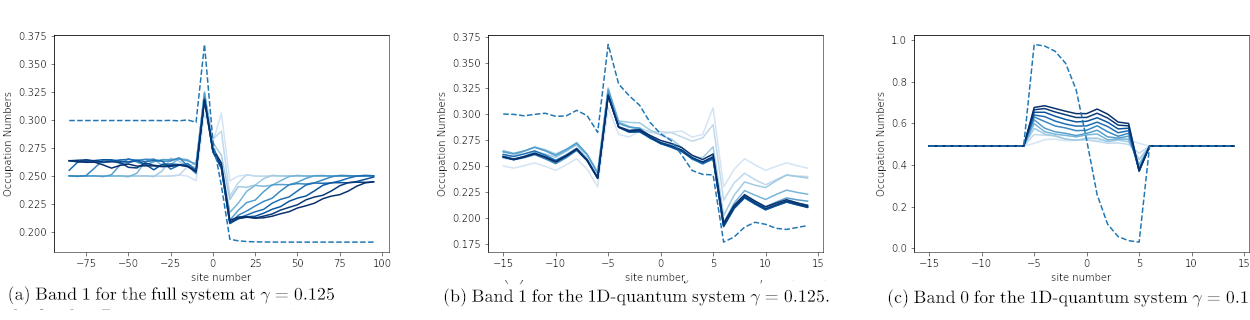}
    \caption{Fermionic occupation number at $\eta=0.2$ for the semi-conductor ($\gamma=0.125$) as a function of the lattice sites for several times. Larger times are encoded via bolder blue lines. The dashed lines represent the expected values for the limit Gibbs state, when the full system is in presence of an electric potential. The 1D quantum system lies between the sites $-5$ and $5$.}
    \label{occupationnumbertime2}
\end{figure}

To complement this discussion, Figs. \ref{occupationnumbertime1} shows the difference between the semiconductor  ($\gamma =0.125$) and conducting cases ($\gamma=0.01$) in terms of time-dependent occupation numbers, at a fixed $\eta=0.1$. In this last case, the second band inside the 1D-quantum system is slightly time dependent, in contrast with the trivial case $\gamma = 0$. But, the effects of the band 0 on currents are negligible while they significantly alter the transport properties at $\gamma =0.125$, as explained above. The saturation phenomenon leading to the semi-conductor like behavior can also be seen by comparing Figs.  \ref{occupationnumbertime1}(a), \ref{occupationnumbertime1}(c), \ref{occupationnumbertime1}(e) for $\gamma =0.125$ and $\eta=0.1$ with Fig. \ref{occupationnumbertime2} for the same $\gamma$ but at $\eta=0.2$, keeping in mind Fig \ref{etatests}(e). 

\subsection{Dependence on the Length of the System}

In 2012, experimental measurements \cite%
{Ohm-exp} of electric resistance of nanowires in Si doped with phosphorus
atoms demonstrate that the macroscopic laws for charge transport are already
accurate at length scales larger than a few nanometers, even at very low
temperature ($4.2~\mathrm{K}$). As a consequence, microscopic (quantum)
effects on charge transport can very rapidly disappear with respect to
growing space scales. Understanding the breakdown of the classical
(macroscopic) conductivity theory at microscopic scales is an important
technological issue, because of the growing need for smaller electronic
components.

From a mathematical perspective, the convergence of the expectations of
microscopic current densities with respect to growing space scales is proven
in \cite{OhmIII,OhmVI}. In \cite{Ansta-LDP,Ansta-LDP2}, for non-interacting
lattice fermions with disorder (one-band, any dimension), it is additionally
proven that quantum uncertainty of microscopic electric current densities
around their (classical) macroscopic value is suppressed, exponentially fast
with respect to the volume of the region of the lattice where an external
electric potential is applied. 

Our model does not have any random external potential, but two bands inside
the system of length $2l$, which corresponds in our unit to
\begin{equation*}
2l\times \mathbf{a}=2l\times 3 \textrm{\AA},
\end{equation*}%
see Section \ref{ss:chooseparams}. Therefore, we
analyze how fast the convergence of the current density converges to a fixed
value.

\begin{figure}
    \centering
    \includegraphics[scale=0.28]{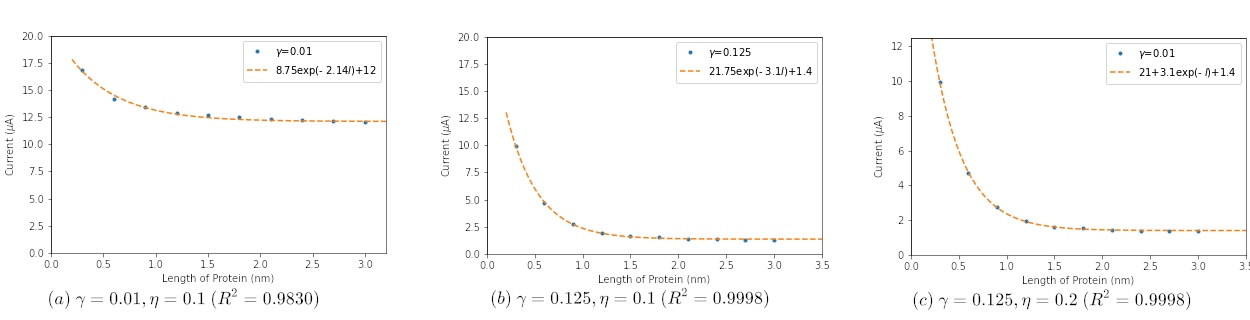}
    \caption{Length dependence for the expectation value of current for the 1D quantum system in the conducting regime for $\eta=0.1$, and semi-conducting regimes at $\eta=0.1$ and $\eta=0.2$ respectively. The dashed line is an exponential fit. Note that we do \emph{not} see a linear decrease of the current with length of the conductor. Here $R^2$ is the usual coefficient of determination.}
    \label{lengthtests}
\end{figure}

In Fig. \ref{lengthtests}, we confirm the exponential convergence of the
current density to a fixed value as
conjectured from \cite{Ansta-LDP,Ansta-LDP2}. In particular,\ in these
numerical experiments, $2\ \mathrm{nm}$ is already a large quantum object
in the sense that the limit point is already reached. Note, finally, that the semiconductor and conductor cases have the same behavior with respect to such convergences.   

\subsection{Dependence on Temperature\label{Dep temp}}

\begin{figure}
    \centering
    \includegraphics[scale=0.45]{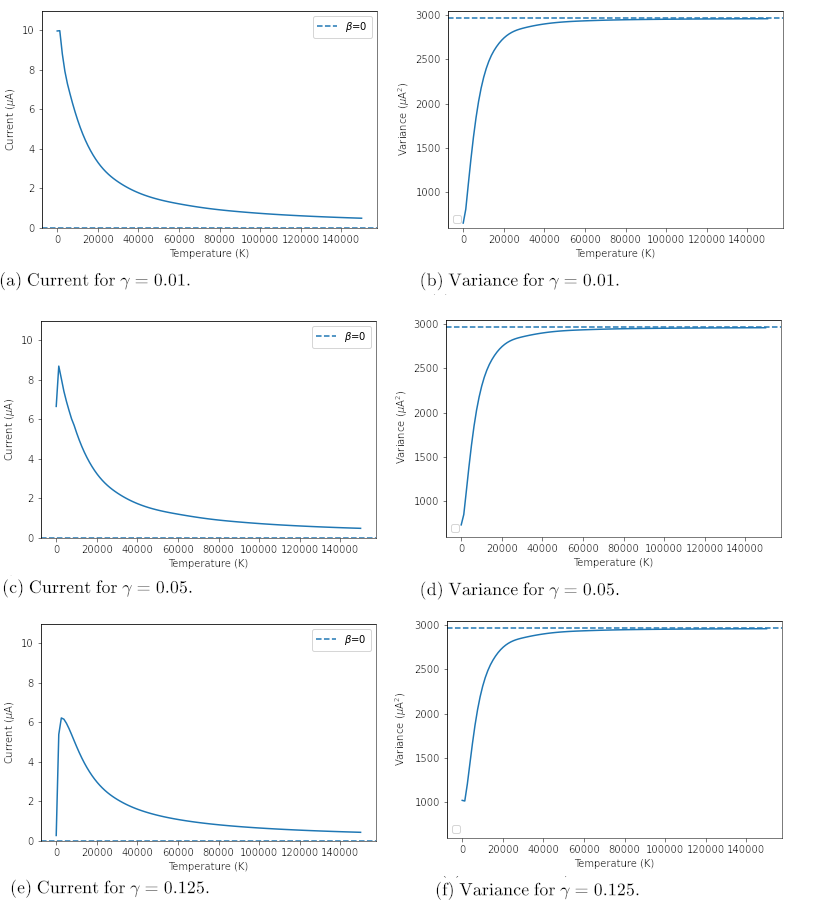}
    \caption{Temperature dependence for the expectation value of current and current variance for the 1D quantum system in the conducting ((a)-(b)), intermediate ((c)-(d)) and semi-conducting ((e)-(f)) regimes at $\eta=0$. The dashed line is the asymptotic for infinite temperature, corresponding to computations at $\beta = 0$.}
    \label{betaasympototicbehavior}
\end{figure}

\begin{figure}
    \centering
    \includegraphics[scale=0.45]{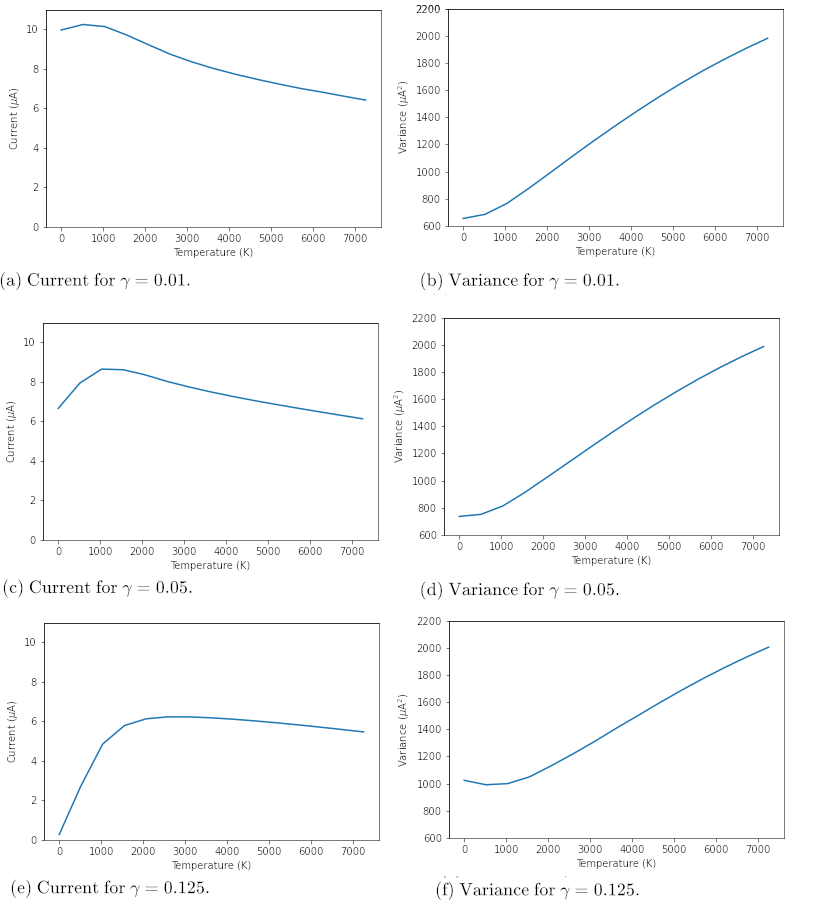}
    \caption{Temperature dependence for the expectation value of current and current variance s for the 1D quantum system in the conducting ((a)-(b)), intermediate ((c)-(d)) and semi-conducting ((e)-(f)) regimes at $\eta=0$ for smaller temperatures.}
    \label{betazoom}
\end{figure}

The energy scale used here are of the order of tenth of volts electronvolts.
This is coherent with the voltage applied in the experiments described in
\cite{zhang2019role}. In this energy scale, a room temperature, i.e., $%
\mathrm{T}\approx 300\mathrm{\ K}$, corresponding to an inverse temperature $%
\beta :=\left( \mathrm{k}_{B}\mathrm{T}\right) ^{-1}\approx 38.7$ $\mathrm{eV%
}^{-1}$, already refers to a very low temperature regime. Consequently, the
transport properties of the system are basically independent of
temperatures below room temperatures.

On the one hand, this is coherent with observations done in \cite%
{temp-independent} on charge transport in some 1D quantum system. On the other
hand, the energy scale used here may be inaccurate in many situations. We
therefore explain the behavior of the model in the high temperature regime,
which basically refers to $\beta \leq 1$, even if it corresponds to
nonphysical temperatures ($\mathrm{T}\geq 11604\mathrm{\ K}$) in our energy
scale. This is performed in Fig. \ref{betaasympototicbehavior}, showing an expected
decrease of currents (equivalently an increase of the electric resistance) as the
temperature increases. This is true for all regimes (conductor and
semi-conductor). In the same way, the variance increases with the
temperature, since thermal fluctuations are of course added to the purely
quantum ones.  See Fig. \ref{betaasympototicbehavior}.

In contrast with the case $\gamma =0 $ with no insulating band 0, the two-band system has very interesting behavior for small temperatures: The current fluctuations are basically increasing, as naively anticipated, except in the semi-conductor regime for small temperatures. See Fig. \ref{betazoom}. Remarkably, the current reaches a maximum value for some non-zero temperature before decreasing to zero in the limit $\beta \rightarrow 0$. See again Figs.  \ref{betaasympototicbehavior} and \ref{betazoom}. It means in this case that the first excited state of the model interestingly favors the existence of currents, as compared with the ground states. It seems that this occurs for all $\gamma \neq 0$ and this effect increases with $\gamma $ to the point that even the current variance starts to become non-monotone. This is a non-trivial information on the spectral properties of the model.    

\section{Conclusion}
A first modelling attempt, based on many-body (fermionic) quantum statistical principles, is given to explain how Ohm's law emerges in long-range charge transport within protein-ligand complexes, the conductances of which have been measured in aqueous solutions and via STM and Pd substrate \cite{zhang2019role}. To this end, we build upon the mathematically rigorous framework (explored by one of the authors and collaborators), which provides proof for the convergence of the expectations of microscopic current densities and also crucially explains how the microscopic (quantum) effects on charge transport vanishes with respect to growing space scales \cite{OhmVI,OhmIII}. Moreover, for non-interacting lattice fermions with disorder (one-band, any dimension), they prove that \textquotedblleft quantum uncertainty of microscopic electric current densities around their (classical) macroscopic value is suppressed exponentially fast with respect to the volume of the region of the lattice where an external electric field is applied" \cite{Ansta-LDP2,Ansta-LDP}.

Ultimately, we show how charge transport in a two-band lattice quantum system leads to the emergence of Ohm-like law, therefore providing a tentative explanation for the experiments in \cite{zhang2019role}. We also show that there is exponentially fast convergence of the expectations of microscopic current densities. This should be related to the existence of non-vanishing current variance of linear response currents, consistent with previous studies under natural conditions on the (random) disorder \cite{Ansta-LDP2,Ansta-LDP}). We expect that this is also true also for several-band models.

Admittedly however, the model's voltage-current curve exhibits a discrepancy of several orders of magnitude when compared to that of the experiment. Specifically, the voltage applied in the experiments described in \cite{zhang2019role} for single perptides is of the order of tenth of volts and the systems remains in the linear response regime. If this is correct, then the energy scales of the model should be of the same order. This yields a quantitative discrepancy since the rigorously computed currents are of the order of $\mu\mathrm{A}$ (see Figs. \ref{etatests}(a)--\ref{etatests}(f)), whereas the measured currents in \cite{zhang2019role} are of the order of $10^{-1}$ $\mathrm{nA}$, $10^{4}$ times smaller. This issue cannot be a priori solved by reducing the hopping strength $\vartheta \geq 0$ between the reservoirs and the band $1$ of the 1D quantum object (1D quantum system). Preliminary numerical computations show that the value of $\vartheta$ has to be extremely small to reduced the current by $4$ orders of magnitude, at the cost of almost isolating the 1D quantum system from the reservoirs and destroying the main transport properties described here.

Since there is no approximations or a priori assumptions on the model, there is probably an issue related to all the energy scales, meaning that the voltage seen by the protein/peptide is far less than tenth of volts. On the other hand, lower energy scales change the meaning of the low temperature regime, which may be in contradiction with previous results \cite{temp-independent}. See discussions in Section \ref{Dep temp}.







Nevertheless, our framework provides a novel approach to study rigorously charge transport (under quantum and thermal fluctuations) and to understand the breakdown of the classical (macroscopic) conductivity theory at microscopic scales. In particular, the theory allows us to explain the emergence of electronic states that are delocalized within a physical system. Hence we provide predictions for both conducting and semiconducting like properties, although semiconductor type characteristics has not yet been observed since only six protein-ligand complexes have been tested so far \cite{zhang2019role} (to our knowledge). Nevertheless, these predictions could be tested in future experiments. The basic nature of the model allows it to be readily adjusted to explore a variety of different phenomena. For instance, an external potential such as 

$$\sum_{x=-l}^{l}v\left( x \right) a_{x,1}^{\ast}a_{x,1},$$
with $v\left( x\right) \in \left[ -1,1\right] $ being some (i.i.d.) random variables (cf. the Anderson model), could have been considered without breaking the quasi-free property of the model. This term can be used to verify and study the linear dependence of the resistivity with respect to the length $l$ of the 1D quantum system (another version of Ohm's law). Note indeed that this property on the resistivity does not hold true for our two-band model (like for the one-band one, i.e., for $\gamma = 0$), since the current does not seem to vanish in the limit of infinite $l$ (Fig. \ref{lengthtests}) while it is expected \cite{Ansta-LDP,Ansta-LDP2} to be equal to the macroscopic one at relatively small lengths $l$.

In the future, it would be interesting to study the emergence of the telegraph noise as observed in the protein-ligand complex experiments (see \cite[Fig. S7]{zhang2019role}). Presently, as far as we know, there is no theory based from first principles of quantum mechanics to explain telegraph noise. Current theories are mainly phenomenological, modelled via a (Markovian continuous-time) stochastic process that jumps discontinuously between two distinct values, as it is shown for
currents in \cite[Fig. S7]{zhang2019role}. The telegraph noise is in particular characterised by a bimodal distribution of currents. Since this can be characterised via some statistical methods involving higher moments (like the kurtosis and skewness) besides the variance, our approach could be used to understand the appearance of the telegraph noise via a purely quantum microscopic theory.

Finally, our code is freely accessible at:
\\
\href{https://github.com/RoisinMary/quasifree_chargetransport.git}{\textrm{https://github.com/RoisinMary/quasifree{\_}chargetransport.git}} and can readily run in a standard computer (e.g. we run it on a MacBook Pro with a 2,4 GHz Quad-Core Intel Core i5-processor).

\section{Acknowledgment}
 
SR, JBB and RB acknowledge support from Ikerbasque (The Basque Foundation for Science), from the Basque Government through the BERC 2022-2025 program and by the Ministry of Science and Innovation: BCAM Severo Ochoa accreditation CEX2021-001142-S / MICIN / AEI / 10.13039/501100011033. SR further acknowledges the PID2023-146683OB-100 funded by MICIU/AEI /10.13039/501100011033 and by ERDF, EU. JBB meanwhile acknowledges of the grant PID2020-112948GB-I00 funded by MCIN/AEI/10.13039/501100011033 and by ``ERDF A way of making Europe", as well as the COST Action CA18232 financed by the European Cooperation in Science and Technology (COST). JUA aknowledges support from the Spanish Government, grants
PID2020-117281GB-I00 and  PID2019-107444GA-I00, partly from European
Regional Development Fund (ERDF), and the Basque Government, grant
IT1483-22.

\appendix

\section{Appendix: Discussion of the Implementation}\label{times}
\begin{figure}
    \centering
    \includegraphics[scale=0.45]{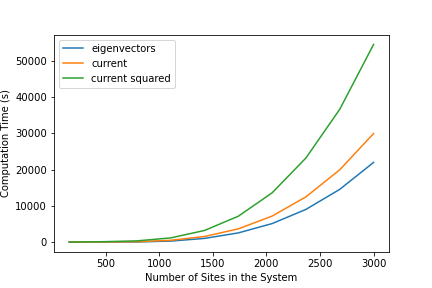}
    \caption{A plot of computation times versus $2(L+1)$, the number of sites in a system. The computation time grows exponentially with increasing $L$.}
    \label{computationalspeed}
\end{figure}
Here computations were performed to find the variance but the code can easily be extended to compute higher moments which could be instrumental in identifying, for example, a bi-modal distribution which is characteristic of telegraph noise. Code automatically implementing the simplifying CAR relations significantly reduces the difficulty of computing higher-order moments. However, because of the increasing number of linear operations needed such calculations can become expensive for $L$ large (see Fig. \ref{computationalspeed}).






\end{document}